\documentstyle[12pt,a4,epsfig]{article}
\begin{document}
\vspace*{-1.0cm}

\begin{center}
{\bf  APPLICATION OF  NEURAL NETWORK FOR PHOTON-HADRON DISCRIMINATION 
 IN A PRESHOWER DETECTOR IN HIGH ENERGY HEAVY ION COLLISIONS}
\vskip 0.9cm
S. Chattopadhyay, Z. Ahammed and Y.P. Viyogi\footnote{Corresponding
author} \\
{\it Variable Energy Cyclotron Centre  \\
1/ AF, Bidhan Nagar, Calcutta-700064 (India) }   \\
\end{center}
\vskip 2.5cm
\begin{abstract}
Using a combination of a preshower detector and a charged particle veto,
it is shown that the neural network method is able to provide satisfactory
discrimination between photons and hadrons  in the case of extremely
high particle density produced in the forward region of heavy ion
collisions at the LHC energy. 
\end{abstract} 

\newpage

 \section{Introduction}
{\indent   Neural network has been extensively used to solve problems of
high
degree of complexity, such as feature classification, image processing,
pattern recognition, combinational optimization etc.
 Many topics encountered in 
high energy physics consist of similar problems of classification
 and pattern recognition and artificial neural
network methods have been successfully applied to solve them
\cite{neu1}.
 One of the specific 
examples is to discriminate different type of shower 
particles in a calorimeter \cite{l3note}. But these applications 
are mostly limited to  situations where the particle multiplicity  is low 
and chances of overlap of hadron and photon showers are negligible.}

	In  heavy ion collisions at ultra-relativistic energies one faces
the problem of  
detection of particles in a very high multiplicity environment.
In particular, for photons, which are conventionally detected using
calorimetric techniques \cite{saphir}, it becomes increasingly difficult
to separate
fully developed showers in the forward pseudorapidity ($\eta$) region
because of high particle density.
In such situations photon studies have been limited to 
the measurement of  spatial distribution using a 
preshower detector
\cite{wa93nim}.
Such a detector consists of a converter (usually 3 $X_0$ thick lead
plates)
 followed by a finely granular sensitive medium  e.g.,
scintillator pads or gas cells. 
The sensitive medium responds to  electromagnetic 
shower
particles generated in the  converter. This
technique considerably reduces the possibility of  overlap 
of showers.

Even with the use of a preshower detector photons need to be discriminated
from hadrons which also produce signal in the sensitive medium. 
Most of the charged
hadrons, while passing through the detector, deposit energy equivalent to
the Minimum Ionising Particle (MIP) in the sensitive medium and are
usually confined to one pad. Photon signal, on the other hand, corresponds
to several MIPs depending on the number of shower particles and spreads to
several cells. After clustering of pad hits,
 hadron rejection is 
achieved by applying a suitable threshold (taken usually equivalent to 3
MIPs)  on the cluster signal. Clusters having signal above threshold are
more likely photon candidates and are referred as "$\gamma$-like"
clusters.

Impurity in the photon sample due to hadron
contamination should be less than a few percent 
above 3 MIP threshold, arising from the tail of the MIP distribution,  if
all hadrons behaved as 
MIPs. However a 
sizeable fraction of hadrons interacts in the converter (3 X$_0$ thick
lead
is equivalent to $\sim$ 10\% of interaction length) and generates
signal which looks like photon shower.  These hits cannot be filtered by
the application of threshold.  Two closely moving hadron tracks may also
mimick a photon signal. Upstream
conversion of photons in the intervening material close to target may lead
to more than one cluster on the detector.  Hadron interactions in the
converter and extra
photon clusters constitute a large contamination in the photon sample. It
is therefore necessary to use new approaches using information which
provide improved discriminatory properties between hadron and photon
signals in a preshower detector.

 Charged Particle Veto (CPV) detector placed in front of an 
electromagnetic
calorimeter has been used in tagging charged hadrons and for separating
photons \cite{saphir}.  Similar arrangement can be attempted with
preshower detector
also.
Both the CPV and the
preshower detector respond to incident charged particles but the CPV, by
its
design, will be insensitive to photons. Thus charged particles can be
vetoed by comparing the signals in the two detectors.
However in a high multiplicity environment difficulties arise in the use
of  proper algorithm for vetoing   because of two reasons:

{\noindent (i) Finite separation between the preshower 
 detector and  the CPV 
and  multiple scattering in the converter destroys  the one-to-one 
correspondence in coordinates between the cells hit by a charged particle
in the two detectors, and}

{\noindent (ii)  closeness of hadron and photon tracks in high particle
density
environment results in  a large probability of 
vetoing   photon clusters, thereby reducing the efficiency of
photon detection.}

In the present article we explore the 
applicability of  neural network (NN) methods with
several inputs representing the discriminatory properties between hadrons
and photons and compare their performances with the results of simpler
vetoing algorithms. We present simulation results for a realistic case of
a preshower detector expected to measure the spatial distribution of
photons in heavy ion collisions
at the LHC energy. Preliminary results have been presented elsewhere
\cite{dae_neu}.

\section{Detector Simulation}

{\indent  The preshower detector under study consists of 3 $X_0$  thick
lead
converter with cells of 
gas proportional chamber as the
sensitive medium. The detector covers typically one unit of pseudorapidity
in the forward cone and is arranged in the form of rectangular matrices of
 gas cells, so that each cell  is
surrounded by eight neighbours in a
3 $\times$ 3 matrix.  Considering
 the estimated particle density in Pb + Pb collisions at the LHC energy,
 we have used square cells  of 1 cm size and 8 mm depth.
 A CPV having the same granularity as the preshower part has  been
placed in front of the converter, the cells being back-to-back with the
preshower
cells. 
The detector is placed at 6 meter from the interaction
point.  No other material has been placed in the intervening space.}

  The response of the preshower detector and the CPV has been simulated
using GEANT3.21  simulation package \cite{geant}. 
For a systematic investigation of the usefulness of vetoing algorithms,
two
different cases have been studied :

(a) single particles : photons and positive pions at specific energies of
1 GeV, 2 GeV, 5 GeV and 10 GeV.

(b) event generator :  particles generated from VENUS 4.12 
event generator \cite{venus} for central (b $<$2 fm) Pb + Pb collisions at
the
LHC energy.

Particle multiplicity in VENUS events
is one of the highest among all the event generators currently being used 
for detector simulation studies at LHC energy \cite{alice}. Hence for
detector design
 it provides one of the worse  case scenarios.

GEANT results in the form of energy deposition in gas cells have been used
for clustering and photon measurement. 
Cells having energy deposition less than that equivalent to 0.2 MIP are
not considered, this level being taken as the detector noise.
No attempt has been made to include
the response of gas proportional chamber. Nearest  neighbour
clustering algorithm has been used to find out photon and hadron clusters.
A fraction of photon and hadron tracks produces more than one cluster.
Clustering of CPV hits has not been attempted.

An efficient photon-hadron discrimination algorithm should be able to
select all the clusters ($N_{cls}^\gamma$) bearing photon identity from
 the total number  of clusters ($N_{cls}$)  found on the
preshower detector.    The discrimination is usually achieved by
computing some observable and applying a threshold. Number of clusters
above such a  pre-selected
discrimination
threshold is called $\gamma$-like clusters
($N_{\gamma-like}$). 
 This contains majority of photons and some
contaminants which reduce the purity of the photon sample. 

For a quantitative description of the effectiveness of these algorithms,  
we define the following two variables :

$\epsilon_s = N^{\gamma,th} _{cls} / N^\gamma _{cls}$ 

$f_p = N^{\gamma,th} _{cls} / N_{\gamma-like}$ 

{\noindent where $\epsilon_s$ is  photon selection
efficiency, $f_p$ is the fractional purity of the photon sample and
 $N^{\gamma,th} _{cls}$
is the number of
photon  clusters above the threshold.}

The photon selection efficiency discussed here is different from the
photon counting efficiency $\epsilon_\gamma$ described in Ref.
\cite{wa93nim}.  $\epsilon_\gamma$ will be smaller than $\epsilon_s$ by a
factor which takes into account the conversion probability of photons in
lead and also the loss (or gain) of clusters due to particular clustering
algorithms used and because of high multiplicity effects.

By adjusting the  discrimination threshold it is  possible to
obtain a highly pure sample
of photons, but the efficiency of selection will be small. On the other
hand if one wants to retain most of the photons in the sample (i.e., high
selection efficiency) then the purity of the sample will be lower. In
practice the tradeoff is dictated by physics considerations, a 
photon sample having at least 80\% purity  being  considered reasonable
for
the study  of
particle production and  fluctuation on event-by-event basis.

\section{Vetoing Algorithms}

\begin{figure}
\centerline{\epsfig{figure=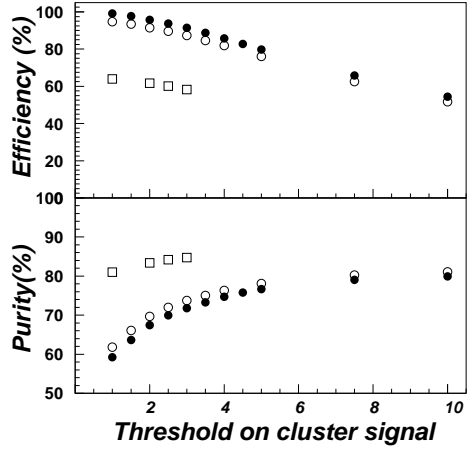,width=8cm}}
\caption{Selection efficiency and purity of photons as a function of
threshold on the cluster signal. Filled circles represent the
results of simple case
without using CPV hits, open circles  and open squares represent 
respectively the results of
 Veto-1 and Veto-9 methods.}
\label{thres}
\end{figure}

{\indent Depending upon the discriminatory properties of different
variables 
obtained  for different types of particles, 
one can use suitable algorithm to separate them. 
In the absence of a charged particle veto, hadrons are filtered using a
threshold on the cluster signal \cite{wa93nim}. Fig. \ref{thres}
shows $\epsilon_s$ and $f_p$ as a function of threshold on cluster signal
for the case of particles from VENUS event generator.
Applicability of this method is severely limited if photon samples with
purity in excess of 70\% are desired, photon selection efficiency falling
sharply from 94\% for 70\% purity to a paltry 55\% for 80\% purity.}

Vetoing of charged particles using information from CPV hits depends
critically on the algorithms selected. Following two algorithms, which 
may be successful in low multiplicity environments, have been tried
here : 

(a) If signal in the CPV cell directly opposite to the cluster maximum in
the preshower part
 is non-zero, then veto the cluster (referred as Veto-1).

(b) If total CPV signal in 3 $\times$ 3 matrix opposite to the cluster
maximum is non-zero then veto the cluster (referred as Veto-9).

Combining the veto and threshold on cluster signal one can select photon
samples with varying efficiency and purity. Results for both the vetoing
methods are also presented in Fig. \ref{thres}.
The Veto-1 method is able to reject only a small number of clusters,
resulting in a relatively high photon selection efficiency at the cost of
purity.
Still this method performs better than simple threshold on the
cluster signal at higher purity, giving 10\% higher selection efficiency
for the
desired 80\% purity.
The Veto-9 method rejects lots of  clusters including those of photons.
Hence even though
purity of $\gamma$-like clusters may be high, efficiency of photon
selection
is always low and never more than 65\%. In this respect Veto-1 method is
more useful.

\section{Neural Network}

{\indent Although the use of a CPV and a  simple vetoing algorithm helps
to improve
the performance of the preshower detector, the  efficiency is far
too less to be of practical use. Hence it is imperative to search for more
intelligent algorithms like neural networks which can utilize maximum
information of the shower
profile and provide improved discrimination of photons from hadrons.}

\subsection{Setup and Training}

{\indent	We setup a standard feed-forward network using upto 21
inputs
derived from  preshower and CPV hits of
  each observed cluster of photons and hadrons.
Two  approaches have been followed for training and testing the network.}

 (A) {\it Hidden  Layer Approach}: This approach has been
used quite extensively in pattern recognition work. Here we use a 3-layer
network with upto 21 inputs in the hidden layer and one in the output
layer.
JETNET 3.4 NN package has been used for this approach \cite{jetnet}. 

 Different algorithms e.g., standard backpropagation algorithm, Manhattan
updating and Langevin method  have been tried for training the network. 
The standard back-propagation algorithm has been found
to be most effective in the present case. The training  parameters used in
this algorithm are :
learning parameter = 0.2, 
momentum ($\alpha$) = 0.05, network temparature = 1.

Training involves the determination of weights by
 minimizing  the error function
   $(F(x) - F_{known})^2$,

{\noindent where $F(x)$ is the response to the inputs $x$ and is also 
a function of the weights applied to the network. The response function used
is the sigmoid function :}

\begin{equation}
F(x)=\frac{1}{1+e^{-x}}
\end{equation}

Target output values ($F_{known}$)
for photon and hadron samples 
used were 1 and 0 respectively.

(B) {\it Functional Expansion Approach:}  In this method, we scale 
each variable ($x_k$)
in the range
(-1, 1), where $k$ varies from one to the number of inputs
 for each cluster  \cite {tariq}.
 It is then expanded using orthonormal functions,

\begin{equation}
f(x_k)=w_{1k}x_k+\sum_{i=1}^{i=n}(w_{2ki}sin(i{\pi}x_k)+w_{3ki}cos(i{\pi}x_k))
\end{equation}

{\noindent where $w$ is a set of weights and $n$ the order of expansion.
We have used
$n$=4 for
the present study. The transformed input quantities are then summed to
form}

\begin{equation}
X=\sum_{k=1}^lf(x_k) + w_0
\end{equation}

{\noindent where $l$ is the total number of inputs.}

The classifier function used here is also a sigmoid function given by,

\begin{equation}
F(X)=\frac{1}{1+e^{-X}}
\end{equation}

The target output values  are the same as for
hidden layer case.
In the present study most of the results are obtained using
approach (A). The effectiveness of approach (B) is described only 
in the context of results for the event
generator case.

\subsection{Inputs}

{\indent Using GEANT simulation results and clustering on the preshower
part as
described in Sec. 2, following variables
 are  extracted for each cluster  to be used as NN input :}

(i) signal strengths in 9 cells 
($ A_1$  to $A_9$ )
around the cluster maximum
of the preshower detector, 

(ii) signal strengths in 9 cells in CPV 
($ A_{10}$ to $A_{18}$)
around the cells situated
opposite to the cluster maximum mentioned in item (i). 
 
(iii) total number of cells affected by the cluster in the preshower part
of the detector ($A_{19}$). 

(iv) total signal strength of the preshower cluster ($A_{20}$) (= sum of
$A_1$ to $A_9$),

(v) total energy deposition in 9  CPV cells around the 
cluster maximum in the preshower part ($A_{21}$) (= sum of $A_{10}$ to
$A_{18}$).

\begin{figure}[h]
{\centerline{\epsfig{figure=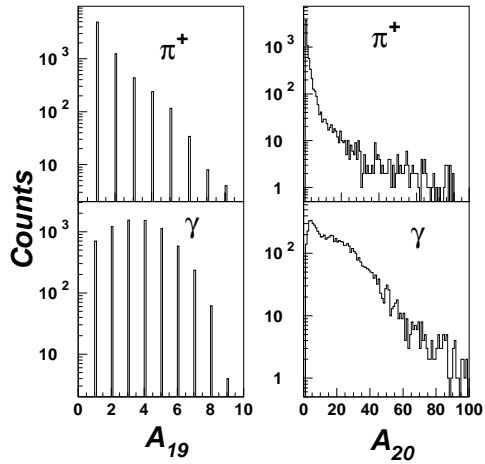,width=8cm}}}
\caption{Input variables
$A_{19}$, $A_{20}$ for 2 GeV
single particles ($\gamma$ and $\pi^+$) illustrating the different
responses for  photon-hadron discrimination.}
\label{input}
\end{figure}

Comparison of all the inputs for photons and hadrons have been made
 to study their discriminatory properties.
 Fig. \ref{input}  displays  inputs
  $A_{19}$ and  $A_{20}$  for photons and for $\pi^+$'s at 2 GeV.
The discriminatory behaviour of  input
$A_{20}$ is  significant. This is the reason that threshold on cluster
signal has been used earlier for hadron rejection in a preshower detector
\cite{wa93nim}.

\subsection{Preprocessing}

{\indent     The  numerical values of the input variables 
 vary widely from each other e.g., the numerical 
value of cluster signal  in eV units in the preshower part ($A_{20}$)
is
about  10000 times higher than the number of cells affected ($A_{19}$).
 Such a large variation leads to the
application of non-uniform weightage while processing through the
network. It is therefore necessary to preprocess the inputs to bring them
in the same scale before applying to the network.}

We have used two preprocessing techniques :

 (i)  a  simple method, in which the inputs are multiplied 
 with suitable factors to keep all of them
in the same scale (0 to 1). The result using this method of
preprocessing has been discussed in Ref. \cite {dae_neu}.

(ii) use of Principal Component Analysis (PCA) technique  \cite{pca}.

 The PCA method  is based on eliminating the
correlation existing among the input variables.
A  matrix is formed having
columns equal to the number of inputs in a cluster and rows equal to the
number of
clusters.
This matrix is then replaced by a matrix of the same dimensions,
where the elements are modified in such a way that the
correlations are lost.

No preprocessing is applied
for the  functional link method, because
the   inputs are automatically scaled into (-1,1) range as the first step 
of processing in this approach.

\section{ Performance of the Network}

{\indent The performance of the network is evaluated by the two
quantities, photon
selection efficiency $\epsilon_s$ and  purity $f_p$ defined
earlier.
We describe the results for different approaches below.}

\subsection{ Hidden Layer Approach}

\subsubsection{Single particle case}

{\indent  Tracks of photons and $\pi^+$'s, 40000 each  
of particular energy, were simulated through GEANT, half of them 
 being used in training the
network and another half  for testing the network.
All the clusters originating from photon
and  hadron tracks are identified as photon and hadron clusters
respectively. 
 For Pb + Pb collisions at the LHC energy the ratio of  
 charged  particle multiplicity and  photon multiplicity is close
to one, 
hence we have studied the properties
of the network for equal population of photons and hadrons.}

\begin{figure}[h]
{\centerline{\epsfig{figure=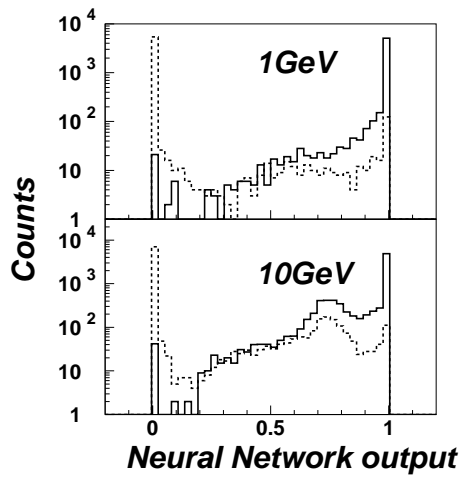,width=8cm}}}
\caption{NN output spectra for the case of single particles 
 at 1 GeV and 10 GeV energies.
 Solid line represents photon clusters and
 the dashed line represents pions.}
\label{spout1}
\end{figure}

Fig. \ref{spout1} shows the NN output spectra for  particles with
energies
of 1 GeV and  10 GeV respectively.
Even though the inputs vary considerably,  the use of
sigmoid function in the network produces  outputs which are 
 peaked around 0 and 1.
The NN output values of clusters can be used for discriminating the two
types
of particles by applying proper threshold.

\begin{figure}[h]
{\centerline{\epsfig{figure=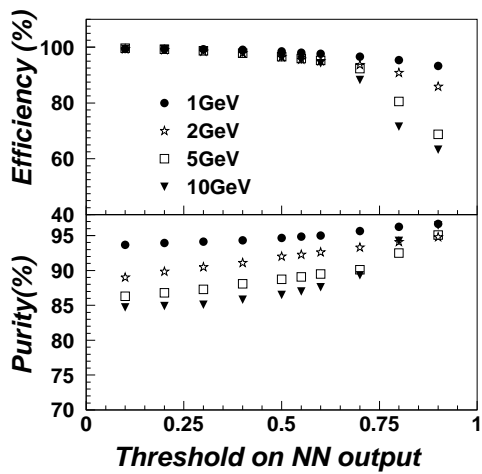,width=8cm}}}
\caption{Efficiency and purity as a function of  threshold on NN output 
in  single particle case at different energies.}
\label{spout2}
\end{figure}

 We have calculated photon selection efficiency and  purity for
different
thresholds applied to the  NN output.
This is shown in  Fig. \ref{spout2}
 for different particle energies.

With  a threshold of 0.55, we observe that
99$\%$ of the input clusters originating from photons can be filtered.
The purity of the detected photon samples  decreases from
 $94\%$ for 1 GeV particles  to $85\%$ for 10 GeV particles. 
At higher energies chances of splitting of clusters increases. Extra
clusters not having proper CPV inputs remain as contaminants, thus
reducing the purity of the photon sample.

\subsubsection{Event generator case}

{\indent GEANT results for particles from the VENUS event generator have
been
processed through the
clustering routine, and the outputs for each cluster 
preprocessed by the PCA technique before using them as the inputs to
the network. All particles in the event including neutral hadrons have
been retained.}

 Because of high particle multiplicity, each event gives a large
number of clusters.
The clusters are
assigned the particle
identity of the track depositing energy on the cluster maxima. 
If both a photon and a hadron hit the same cell, we consider it as a
photon hit.
This treatment is different
from the method adopted in case of single particles, where the clusters
are tagged according to the identity of the incident track only. 
All particles other than photons are considered as  contaminants in
the photon sample. If the photon
track is shifted considerably (0.1 unit in $\eta$ or $10^0$ in azimuthal
angle) from the original direction while forming the cluster, then we
tag the cluster as a contaminant.  This constraint is imposed from the
consideration of acceptable distortion in the measured pseudorapidity
distribution of photons.

For the present study we have used  training sample of about 60000
clusters obtained from 10 events and 
 another set of about 90000 clusters from 15 events for testing the
performance of the network.

\begin{figure}[h]
{\centerline{\epsfig{figure=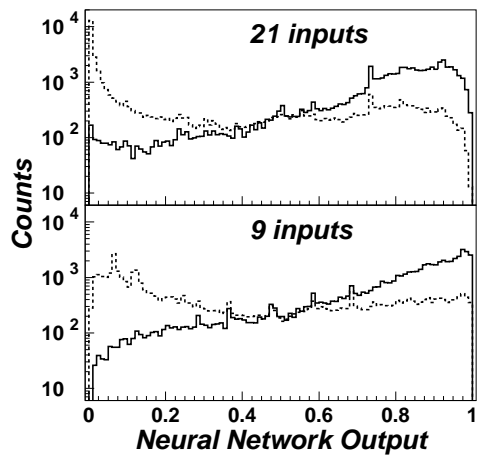,width=8cm}}}
\caption{NN output spectra, for particle clusters 
 from VENUS event generator,  obtained by using 9  and 21 inputs
in the hidden layer approach after  PCA preprocessing.
 Solid line represents photon clusters and 
 the dashed line represents contaminants.}
\label{parinput9}
\end{figure}

We have also studied the effect of different sets of inputs on the
performance of the
network. Results have been obtained for 
9 ($A_1$ to $A_9$),
 18 ($A_1$ to $A_{18}$)
 and 21 ($A_1$ to $A_{21}$) input cases.
NN output spectra for 9- and 21- inputs  are shown in Fig.
\ref{parinput9}.
It is clearly seen  that the discrimination is better with 21 inputs.
Detailed comparison of the performance in various cases is discussed in
the next section.

\subsection{ Functional Link Method}

\begin{figure}[h]
{\centerline{\epsfig{figure=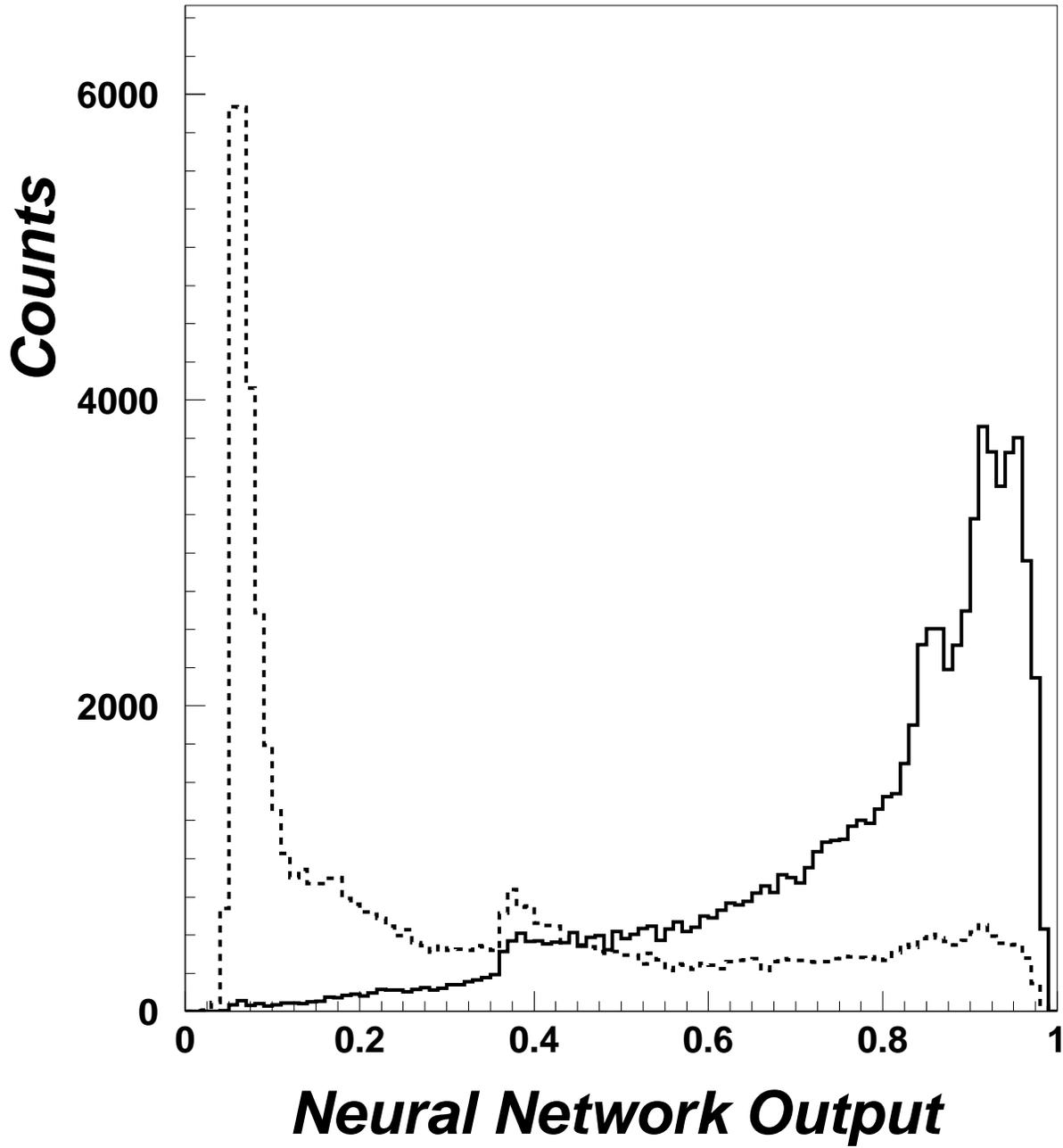}}}
\caption{NN output spectra  in the functional link method for  particle
clusters  from VENUS event generator.
 Solid line represents photon clusters and 
 the dashed line represents contaminants.}
\label{func}
\end{figure}

{\indent Performance of the functional link method has been studied for
particles
from the event generator using 21 inputs.
Fig. \ref{func} shows the output spectrum of the  network.
Appearance of two distinct
peaks in the network output spectrum suggests  that the output can be 
satisfactorily used for hadron-photon discrimination.
 The efficiency and purity values are quite close to those
 obtained from the hidden layer approach with 21 inputs.}

\section{Discussions and Summary}

\begin{figure}[h]
{\centerline{\epsfig{figure=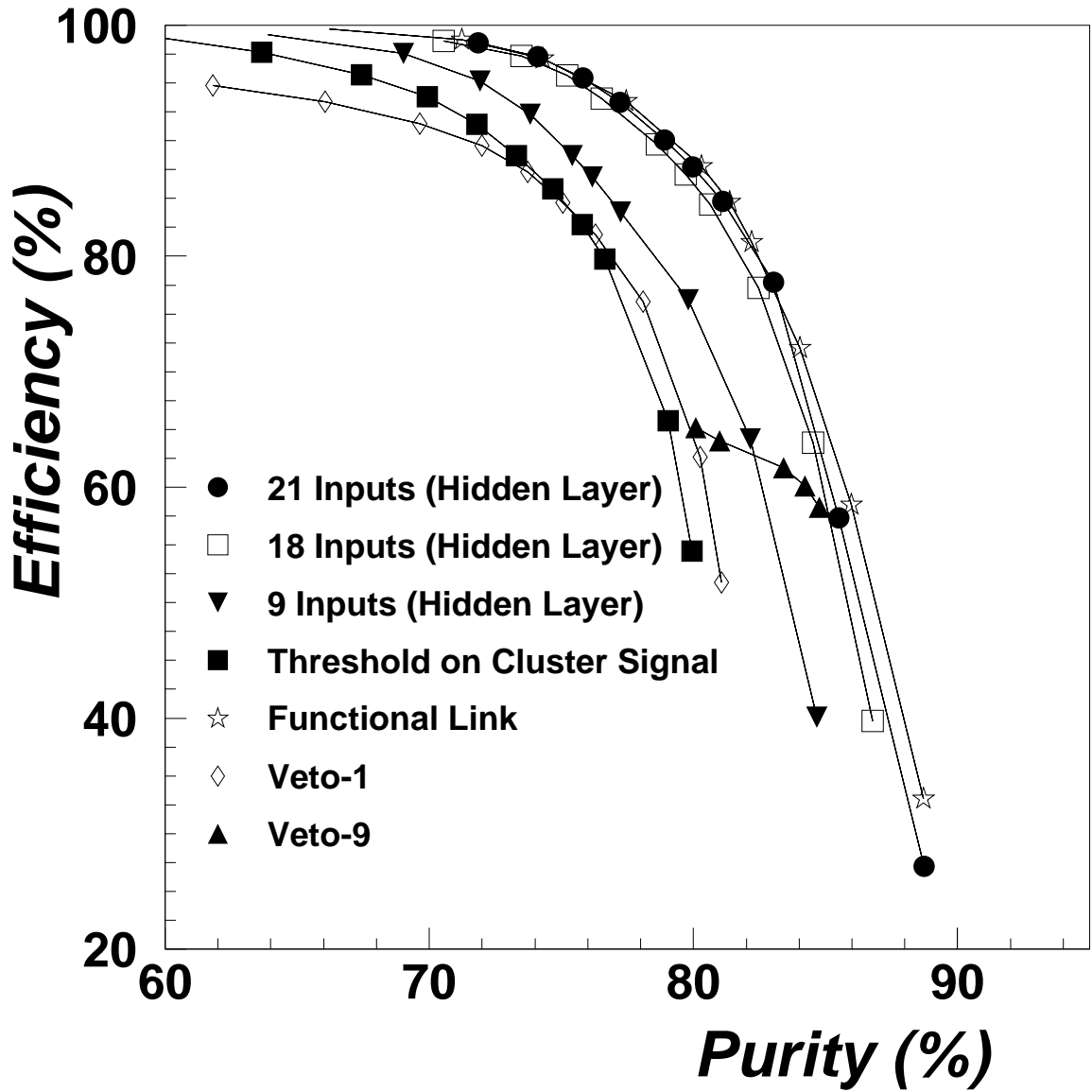}}}
\caption{Photon selection efficiency and purity obtained after applying 
different  discrimination  thresholds  in the
case of particles from the event generator. 
 For hidden layer NN method, the results for 
 9, 18 and 21 inputs are separately shown. The continuous lines are drawn 
only for joining the points.}
\label{cutevt}
\end{figure}

{\indent Photon-hadron discrimination thresholds (either the cluster
signals or the
NN output values) can be adjusted to provide various pairs of values of
$\epsilon_s$ and $f_p$. Fig. \ref{cutevt}  summarizes the results of all
the methods investigated in the present work.}

In general photon selection efficiency drops sharply with increasing
demand on purity of the sample in excess of 70\%. At any given purity
level, the efficiencies for various methods vary approximately according
to the following trend :

cluster signal $<$ Veto-1 $<$ 9-input NN $<$ 18-input NN $<$ 21-input NN

Veto-9 method is not at all suitable in high multiplicity environments.
The NN method using detailed information on the shower profile (9-input
case) even without the use of CPV can provide much better hadron rejection
than the method using only the threshold on cluster signal. Thus for 80\%
purity 9-input NN
method gives a selection efficiency of 76\% compared to 55\%  for the
cluster threshold method. Considering the three NN cases (9-, 18- and 21-
inputs) in the hidden layer approach, it is found that the performance
improves as more inputs are included. However improvement from 18-input
case to 21-input case is marginal, most likely because inputs $A_{20}$ and
$A_{21}$ are correlated with other inputs.

For the desired 80\% purity of photon sample we obtain selection
efficiency of 88\% using 21-input NN method in both the approaches. This
is a very significant improvement, particularly considering the fact
that with only 55\% photon selection efficiency (as obtained by threshold
on the cluster signal) event--by--event physics might not have been
possible. 

Now that a trail has been found, one needs to chart the course further.
The functional link method is computationally faster and simpler to
implement. One needs to study this in more detail as one will be required
to process large volumes of data in LHC experiments.
Studies are in progress on the use of improved clustering routine to
reduce the chances of splitting of
clusters and better discriminatory variables as NN inputs. An important
aspect under study is the effect of increasing the CPV cell size.
Occupancy of CPV cells is typically a factor of 4 lower than the preshower
part. If CPV cell size can be increased without affecting the performance
of the network, this will considerably reduce the cost of the detector.
Results of these investigations will be reported in future publications.

In summary
  neural network technique has been used to separate photon clusters
from the mixture of photon and hadron clusters in a preshower detector
in high energy heavy ion collisions. 
The results have been obtained from the simulated output using a
preshower configuration including a charge particle veto. The difference
in the profile of the energy deposition in the detector by photons and
by hadrons are used for the discrimination. Two techniques of training,
one using standard 3-layer network architecture and the other using the
functional transformation of the inputs give similar results. 
The method is particularly useful in providing filtered photon samples of
high purity.
For photon samples with 80\% purity,  selection efficiency  close
to 90\% can be easily obtained.
 This is almost 30\% higher than those obtained by
applying threshold on the cluster signal and simple vetoing algorithms.


{\bf Acknowledgements}

The authors would like to thank  Tariq Aziz of TIFR,
 Mumbai and  Tapan Nayak,  Murthy S. Ganti and
M. R. Dutta Mazumdar of VECC, Calcutta for useful discussions.

\end{document}